\documentclass[prl,aps,floats,twocolumn]{revtex4}

\usepackage[dvips]{graphicx}
\usepackage{amssymb} 
\usepackage{amsmath}
\usepackage{ctable}

\begin{document}

\title{Cancelling the vacuum energy and Weyl anomaly in the standard model \\ with dimension-zero scalar fields}

\author{Latham Boyle$^1$ and Neil Turok$^{1,2}$} 

\affiliation{$^{1}$Perimeter Institute for Theoretical Physics, Waterloo, Ontario, Canada, N2L 2Y5 \\
$^{2}$Higgs Centre for Theoretical Physics, University of Edinburgh, Edinburgh, Scotland, EH8 9YL}

\date{October 2022}

\begin{abstract}

The standard model is a remarkably consistent and complete quantum field theory but its coupling to gravity and the Higgs field remain problematic, as reflected in the cosmological constant problem, the Weyl anomaly, and the hierarchy puzzle.  We point out that 36 conformally-coupled dimension-zero scalar fields can simultaneously cancel the vacuum energy and both terms in the Weyl anomaly, if the Higgs field is emergent.  The cancellation is highly non-trivial: given the standard model gauge group $SU(3)\times SU(2)\times U(1)$, it requires precisely $48$ Weyl fermions, {\it i.e.}, three generations of standard model fermions, including right-handed neutrinos.  Due to a large additional gauge symmetry, the new scalars contribute no new local degrees of freedom or particle states.  Their only physical state is their vacuum state, in which they possess a scale invariant power spectrum extending to long wavelengths. This suggests a new explanation for the primordial scalar perturbations in cosmology, not requiring inflation. We also discuss how the Higgs field might emerge as a composite object. 

\end{abstract}

\maketitle

\section{Introduction}

The standard model is a highly successful description of particle physics \cite{PeskinSchroeder}.  A remarkable feature is the nontrivial cancellation of all local chiral anomalies (both gauge and gravitational), which severely constrains the matter content \cite{Minahan:1989vd}.  Setting the Higgs field aside momentarily, the gauge/fermion sector is classically Weyl invariant.  As has long been noted \cite{Weyl, Brans:1961sx, Dicke:1961gz}, Weyl invariance -- the freedom to change the unit of length locally -- is a natural extension of diffeomorphism and local Lorentz invariance with strong theoretical motivations (also see {\it e.g.}\ \cite{tHooft:2011aa, tHooft:2014swy, tHooft:2016uxd}). 

However, when we attempt to embed the standard model in curved spacetime, two famous problems arise: the vacuum energy is badly divergent, the observed value being miniscule compared to the natural theoretical expectation (the cosmological constant problem \cite{Weinberg:1988cp}); and the model's Weyl symmetry becomes anomalous (even without the Higgs) \cite{Duff:1993wm}.  In this paper, we point out that one change to the standard model, namely the inclusion of dimension-zero Weyl-invariant scalar fields, simultaneously and uniquely resolves both of these issues through a non-trivial cancellation of the vacuum energy and both parts of the Weyl anomaly (the $a$-term and $c$-term).  

Moreover, we note two remarkable side-effects: (i) the cancellation {\it requires} exactly three generations of standard model fermions (including right-handed neutrinos); and (ii) these new scalar fields have no local degrees of freedom or particle states, but they have a scale invariant power spectrum of vacuum fluctuations that, in the context of our recent proposal \cite{Boyle:2018tzc, Boyle:2018rgh, Boyle:2021jej, Turok:2022fgq, Boyle:2022lcq, Boyle:2022lyw}, could explain the observed primordial scalar perturbations in cosmology. 

This striking set of consequences, ranging from the far UV to the far IR, make the idea intriguing to us. However, we emphasize that this paper is only an initial exploration at the level of free 
quantum fields on a curved spacetime background \cite{BirrellDavies, MukhanovWinitzki}.  The vacuum energy and the Weyl anomaly, as well as the hoped-for emergence of a composite Higgs, are strongly affected by interactions. We comment on these matters in the Discussion. 

\section{Dimension-zero scalars}

Recall that an {\it ordinary} conformally-coupled scalar (with mass dimension one) is described by the action
\begin{equation}
  \label{S_2}
  S_{2}[\varphi]=\frac{1}{2}\int d^{4}x\sqrt{g}\,\varphi\,\Delta_{2}\,\varphi
\end{equation}
where $\Delta_{2}$ is the unique conformally-invariant second-order differential operator \cite{Penrose}
\begin{equation}
  \label{Delta_2}
  \Delta_{2}=-\Box+\frac{1}{6}R\qquad\qquad(\Box=g^{\mu\nu}\nabla_{\mu}\nabla_{\nu}).
\end{equation}
This action is invariant under the Weyl transformation $g_{\mu\nu}(x)\to\Omega^{2}(x) g_{\mu\nu}(x)$, $\varphi(x)\to\Omega^{-1}(x)\varphi(x)$.

By comparison, a {\it dimension-zero} conformally-coupled scalar is described by the action
\begin{equation}
  \label{S_4}
  S_{4}[\varphi]=\pm\frac{1}{2}\int d^{4}x\sqrt{\pm g}\,\varphi\,\Delta_{4}\,\varphi
\end{equation}
where the upper (resp. lower) sign applies in Euclidean (resp. Lorentzian) signature: the Euclidean sign comes from demanding positivity of the Euclidean action, and the Lorentzian sign follows from Wick rotation.  And $\Delta_{4}$ is the unique conformally invariant fourth order differential operator \cite{Fradkin:1981jc}
\begin{equation}
  \label{Delta_4}
  \Delta_{4}= \Box^{2}+2R^{\mu\nu}\nabla_{\mu}\nabla_{\nu}-\frac{2}{3}R\,\Box+\frac{1}{3}(\nabla^{\mu}R)\nabla_{\mu}\,.
\end{equation}
This action is invariant under the Weyl transformation $g_{\mu\nu}(x)\to\Omega^{2}(x) g_{\mu\nu}(x)$, $\varphi(x)\to\Omega^{0}(x)\varphi(x)$. Most of this paper deals with the local properties of the theory (\ref{S_4}); in the Discussion we turn to its effects in cosmology. 

In flat spacetime, the action (\ref{S_4}) reduces to
\begin{equation}
  \label{S_4_flat}
  S_{4}=\pm\frac{1}{2}\int d^{4}x\varphi\Box^{2}\varphi=\pm \frac{1}{2}\int d^{4} x(\Box\varphi)^{2}.
\end{equation}
Since $S_{4}$ is fourth-order in derivatives, one should worry about the possibility of ghosts.  However, in addition to its Weyl symmetry, as Ref.~\cite{Bogoliubov} clarified, the action (\ref{S_4_flat}) is invariant under local gauge transformations of the form $\varphi\to\varphi+\chi$ where $\chi$ is any harmonic function: $\Box \chi=0$.  Once this symmetry is properly taken into account, the theory turns out to have no local degrees of freedom at all. All local, gauge invariant operators commute hence the theory has only a single physical state -- the ground state or vacuum (see Sec.~10.2.A in Ref.~\cite{Bogoliubov} for a careful derivation, and \cite{Bateman} for further developments).  As explained in \cite{Hawking:2001yt}, the ground state wavefunction $\Psi$ may be explicitly evaluated as $\Psi(\varphi_{0},\dot{\varphi}_{0})={\rm e}^{-S_{4}[\varphi]}$ where $S_{4}[\varphi]$ is the Euclidean action evaluated on the classical solution $\varphi$ that vanishes in the Euclidean future $t_{E}\to\infty$, and matches the specified values $(\varphi_{0},\dot{\varphi}_{0})$ at $t_{E}=0$.

\section{Vacuum energy}  

Let's see how the ground-state energy of a dimension zero scalar is determined.  As a warm-up, consider an ordinary scalar field with Euclidean Green's function 
\begin{equation}
  \label{G_fourier_decompose}
  G(x,x')\equiv\langle\varphi(x)\varphi(x')\rangle=\!\int\!\frac{d^{3}k}{\!(2\pi)^{3}}{\rm e}^{i{\bf k}\cdot({\bf x}-{\bf x}')}G_{{\bf k}}(\tau,\tau'),
\end{equation}
with $x\equiv (\tau,{\bf x})$. $G_{\bf k}(\tau,\tau')$ obeys the classical equation of motion with a delta-function source
\begin{equation}
  (-\partial_{\tau}^{2}+{\bf k}^{2})G_{{\bf k}}(\tau,\tau')=\hbar\, \delta(\tau-\tau').
\end{equation}
The solution which is regular as $|\tau-\tau'|\to \infty$ is $G_{k}(\tau,\tau')=(\hbar/(2 k)){\rm e}^{-k|\tau-\tau'|}$.  From (\ref{G_fourier_decompose}), we calculate the expectation value of the Hamiltonian $H=\frac{1}{2}[-\varphi_{\!,\tau}^{2}+(\nabla\varphi)^{2}]$ to be
\begin{equation}
  \langle H\rangle\!=\![(-\partial_{\tau}\partial_{\tau'}\!+\!\nabla_{{\bf x}}\nabla_{{\bf x}'})G(x,x')]\Big|_{x=x'}\!\!=\!\!\int\!\!\frac{d^{3}k}{(2\pi)^{3}}\frac{\hbar k}{2},
\end{equation}
the standard result.  

Now consider the dimension-zero scalar field described by (\ref{S_4_flat}).  Once again, we decompose the Euclidean Green's function into spatial 
Fourier modes. $G_{\bf k}(\tau,\tau')$ then obeys 
\begin{equation}
  (\partial_{\tau}^{2}-{\bf k}^{2})^{2}G_{{\bf k}}(\tau,\tau')=\hbar \, \delta(\tau-\tau'),
\end{equation}
whose solution, regular as $|\tau-\tau'|\to\infty$, is
\begin{equation}
  \label{Gk_4}
  G_{k}(\tau,\tau')=\hbar \left(\frac{1}{4k^{3}}+\frac{|\tau-\tau'|}{4k^{2}}\right){\rm e}^{-k|\tau-\tau'|}.
\end{equation}
From (\ref{S_4_flat}), the Hamiltonian (or energy) density is
\begin{equation}
  H=\frac{1}{2} [-(\partial_{\tau}^{2}\varphi)^{2}+2(\partial_{\tau}\varphi)\Box(\partial_{\tau}\varphi)+(\nabla^{2}\varphi)^{2}],
\end{equation}
thus
\begin{eqnarray}
  \langle H\rangle&\!=\!&\big[(-\partial_{\tau}^{2}\partial_{\tau'}^{2}+2\partial_{\tau}\Box_{x'}\partial_{\tau'}+\nabla_{{\bf x}}^{2}\nabla_{{\bf x}'}^{2})G(x,x')\big]\Big|_{x=x'}
  \nonumber\\
  &\!=\!&\int\frac{d^{3}k}{(2\pi)^{3}}(\hbar{k}).
\end{eqnarray}
We see each Fourier mode contributes $\hbar k$ to the vacuum energy, twice the amount for an ordinary scalar, just as one might expect from its naive degrees of freedom~\cite{Fradkin:1983tg}.  

Putting these results together with the classic results for fermions and vector fields, and restoring $\hbar$, we find that the vacuum energy per Fourier mode is given by
\begin{equation}
  \label{E_k}
  E_{{\bf k}}^{(0)}=\frac{\hbar k}{2}[n_{0}-2n_{1/2}+2n_{1}+2n_{0}'],
\end{equation}
where $n_{0}$ and $n_{0}'$ are, respectively, the number of dimension-one and dimension-zero real scalar fields (spin 0), described by the actions (\ref{S_2}) and (\ref{S_4}), $n_{1/2}$ is the number of Weyl or Majorana spinor fields (spin 1/2), and $n_{1}$ is the number of real vector fields (spin 1). As the superscript $(0)$ emphasizes, this is the lowest order result. 

\section{The Weyl anomaly}

When fields are quantized in a curved spacetime, additional divergences arise. In particular, the trace of the stress-energy tensor, which is zero in a Weyl-invariant classical theory, can acquire a finite, state-independent expectation value even after renormalization. The trace measures the failure of Weyl invariance in the quantum theory. It takes the general form
\begin{equation}
  \langle T^{\mu}_{\mu}\rangle=c C^{2}-a E+\xi \Box R\,,
\end{equation}
Here $C^{2}=R^{\alpha\beta\gamma\delta}R_{\alpha\beta\gamma\delta}-2R^{\alpha\beta}R_{\alpha\beta}+\frac{1}{3}R^{2}$ is the square of the Weyl curvature and $E=R^{\alpha\beta\gamma\delta}R_{\alpha\beta\gamma\delta}-4R^{\alpha\beta}R_{\alpha\beta}+R^{2}$ is the Euler (or Gauss-Bonnet) density.  For a QFT on a curved background spacetime, the coefficients $a$ and $c$ are given at one-loop by the expressions \cite{Capper:1974ed, Capper:1973mv, Capper:1974ic, Deser:1974cz, Christensen:1977jc, BunchDavies, Brown:1976wc, Brown:1977pq, Dowker:1976zf, Duff:1977ay, Duff:1993wm}
\begin{subequations}
  \label{ac}
  \begin{eqnarray}
    \label{a}
    a&=&\frac{1}{360(4\pi)^{2}}[n_{0}+\frac{\!11\!}{2}n_{1/2}+62 n_{1}]\qquad  \\
    \label{c}
    c&=&\frac{1}{120(4\pi)^{2}}[n_{0}+\,3\, n_{1/2}+12 n_{1}] 
  \end{eqnarray}
\end{subequations}
where $n_{0}$, $n_{1/2}$, and $n_{1}$ are defined as in Eq.~(\ref{E_k}).  We can ignore the $\xi\Box R$ term, since it can be removed by adding a local counterterm ($\propto R^{2}$) to the Lagrangian.

Thus, in a field theory like the standard model, built from ordinary scalars, Weyl spinors and vectors, all terms on the right-hand side of Eqs.~(\ref{a}, \ref{c}) are positive, so neither of the coefficients $a$ or $c$ can cancel.  

Gravitino (spin 3/2) fields famously contribute negatively to both the $a$ and $c$ coefficient so that, in certain special $N\geq5$ supergravity theories, the $a$ and $c$ coefficients both vanish \cite{Christensen:1978gi, Christensen:1978md, Hawking:1978ghb, Yoneya:1978gq, Townsend:1979js, Kallosh:1979pd, Nicolai:1980km, Nicolai:1980td, Fradkin:1983tg, Gibbons:1984dg, Duff:1982yw, Meissner:2016onk}. The cancellation of Weyl anomalies in these theories is associated with their improved behavior in the UV (fewer divergences) compared to general relativity. 

It is less well known that dimension-zero scalars governed by $S_{4}$ {\it also} contribute negatively to both the $a$ and $c$ coefficients and thus provide an alternative route to cancelling the Weyl anomaly.  Including these contributions, (\ref{a}, \ref{c}) become \cite{Fradkin:1983tg, Gilkey, Riegert:1984kt, Gusynin:1989ky}
\begin{subequations}
  \label{ac_new}
  \begin{eqnarray}
    \label{a_new} 
    a\!&\!=\!&\!\frac{1}{360(4\pi)^{2}}[n_{0}^{}\!+\!\frac{\!11\!}{2}n_{1/2}^{}\!+\!62 n_{1}^{}\!-\!28n_{0}']  \\
    \label{c_new} 
    c\!&\!=\!&\!\frac{1}{120(4\pi)^{2}}[n_{0}^{}\!+\!\,3\, n_{1/2}^{}\!+\!12 n_{1}^{}\!-\,8\,n_{0}'] \quad\;\;
  \end{eqnarray}
\end{subequations}
where $n_{0}$, $n_{1/2}$, $n_{1}$ and $n_{0}'$ are defined as in Eq.~(\ref{E_k}).

\section{Application to the Standard Model}

Now consider the standard model of particle physics, with a right-handed neutrino in each generation, in a curved background spacetime.  There are $n_{0}=4$ ordinary, real scalars in the usual complex Higgs doublet, $n_{1/2}=3\times16=48$ Weyl spinors (16 per generation), and $n_{1}=8+3+1=12$ $SU(3)\times SU(2)\times U(1)$ gauge fields.  In this case, neither $a$ (\ref{a_new}) or $c$ (\ref{c_new}) cancels, and including additional dimension zero scalars (with $n_{0}'$ an integer) is no help.

However, suppose for a moment that the Higgs field is not actually a fundamental field, but rather a composite or emergent field, so we can drop its contributions to Eqs.~(\ref{E_k}, \ref{a_new}, \ref{c_new}) by setting $n_{0}=0$.  We now notice that, if we introduce $n_{0}'=36$ dimension-zero scalars, the coefficient $a$ (\ref{a_new}) vanishes.   But then we are surprised to find that $c$ (\ref{c_new}) and $E_{k}^{(0)}$ (\ref{E_k}) -- completely independent linear combinations -- vanish as well!  In other words, once we introduce $n_{0}'=36$ dimension-zero scalars to cancel the $a$ part of the Weyl anomaly, we find that the {\it full} Weyl anomaly as well as the vacuum energy cancel for free.

Let us rephrase this result.  We can regard the simultaneous vanishing of the vacuum energy and the Weyl anomaly -- {\it i.e.},\ of $E_{k}^{(0)}$ (\ref{E_k}), $a$ (\ref{a_new}) and $c$ (\ref{c_new})  -- as three equations for the four unknowns $\{n_{0}, n_{1/2}, n_{1},n_{0}'\}$.  The unique solution is 
\begin{equation}
  \label{gen_soln}
  n_{1/2}=4 n_{1}, \quad n_{0}'=3 n_{1}, \quad n_{0}=0.
\end{equation}

We emphasize two striking points about the result (\ref{gen_soln}):  

(i) First, it says that, given the standard model gauge group $SU(3)\times SU(2)\times U(1)$ (so that $n_{1}=12$), simultaneous cancellation of the vacuum energy and Weyl anomaly requires precisely four times as many Weyl fermions, {\it i.e.}, $n_{1/2}=48=3\times16$, or three standard model generations. 

(ii) Moreover, cancellation requires that $n_{0}$ vanishes, so that the Higgs field does not contribute to Eqs.~(\ref{E_k}, \ref{a_new}, \ref{c_new}).  Over the years, the conundra raised by the Higgs field (including the vast hierarchy between the Higgs and Planck masses) have led many physicists to speculate that it might be emergent or composite.  If our results are indeed a clue about the real world, they would appear to support this speculation.

\section{Scale invariant fluctuations}

In previous sections, we described the short distance properties of dimension zero scalars and how these might help with consistently coupling the standard model to gravity. In this section, we consider the opposite, extreme infrared regime. From (\ref{G_fourier_decompose}) and (\ref{Gk_4}) we note that, unlike ordinary scalars, these new dimension-zero scalars automatically possess a scale invariant power spectrum after analytic continuation to Minkowski spacetime: at equal times,
\begin{equation}
  \langle\varphi(t,{\bf x})\varphi(t,{\bf x}')\rangle=\int\frac{d^{3}k}{(2\pi)^{3}}{\rm e}^{i{\bf k}\cdot({\bf x}-{\bf x}')}\frac{1}{4k^{3}}.
\end{equation}
In a flat FRW spacetime, when the line element is expressed in conformal time and comoving spatial coordinates, $ds^2=a(t)^2(dt^2-d\bf{x}^2)$,  as a consequence of conformal invariance the scale factor does not appear in the action (\ref{S_4}). Hence, exactly the same calculation applies, and shows that the spectrum of vacuum fluctuations in $\varphi$ is scale invariant in the cosmological sense. 

Recall that we introduced dimension zero fields $\varphi$ in order to cancel the trace anomaly and restore Weyl invariance. So far, we only dealt with the free field contributions, at lowest order. However, as is well known, interactions give rise to additional contributions to the trace anomaly, even in flat spacetime (for a historical account see Ref.~\cite{Hill:2005wg}). In gauge theories the trace anomaly receives contributions proportional to the beta function times the Lagrangian. It is natural to attempt to cancel these contributions with terms involving dimension zero fields. In particular, one can add a classically non-Weyl invariant term $\int d^4 x \sqrt{-g} {1\over 4} (E-{2\over 3} \Box R) \varphi$, which contributes a term $\Delta_{4}\,\varphi$ to the trace (see, {\it e.g.}, \cite{Mottola:2016mpl}). A quantum trace contribution $\Delta T^\lambda_\lambda$ can be cancelled 
by adding a linear coupling $\int d^4 x \sqrt{-g} \varphi \Delta T^\lambda_\lambda$ to the action, so that from $\varphi$'s classical equation of motion,  $\Delta_{4}\,\varphi$  receives an extra term $\propto \Delta T^\lambda_\lambda$. In this way, a classical contribution from $\varphi$ cancels a quantum contribution from running couplings, restoring Weyl symmetry.  (In the context of sigma models in string theory, a similar two dimensional mechanism is well-known~\cite{Callan:1985ia,Thorlacius}). The terms linear in $\varphi$ contribute to the stress tensor and the Einstein equations, so that quantum fluctuations of $\varphi$ generate curvature perturbations in the metric. In this way, the dimension zero fields can give rise to a scale invariant spectrum of scalar density perturbations (but no tensors), without inflation. This result fits very neatly with the two-sheeted CPT-symmetric universe proposed in \cite{Boyle:2018tzc, Boyle:2018rgh, Boyle:2021jej,Turok:2022fgq,Boyle:2022lcq} and will be detailed in a forthcoming paper~\cite{perts}.  

In our discussion above, we emphasized that the action (\ref{S_4_flat}) for dimension zero fields in flat spacetime is invariant under the local gauge symmetry $\varphi\rightarrow \varphi+\chi$  where $\chi$ is any harmonic function of the spacetime coordinates. This gauge symmetry was helpful in seeing that the quantum theory defined by (\ref{S_4_flat}), despite its involving four derivatives, is free of ghost instabilities and in fact possesses a unique physical state namely the vacuum. When generalizing this result to FRW cosmology,  note that all such cosmologies are locally conformally flat (see, {\it e.g.}, Ref.~\cite{Penrose:1986ca}). Hence, in suitable coordinates, the action again reduces to  (\ref{S_4_flat}) and possesses the same local gauge symmetry. Again, all local, gauge invariant operators commute so there is a unique physical state. 

\section{Discussion}

In summary, we have pointed out that a set of 36 conformally-coupled dimension-zero scalar fields:
\begin{enumerate}
\item cancels the vacuum energy,
\item cancels both terms in the Weyl anomaly,
\item does so only for precisely three generations of standard model fermion, 
\item provides a new source of primordial, nearly scale invariant scalar perturbations, and
\item achieves all this without introducing any new local degrees of freedom or particles.
\end{enumerate}
Using dimension zero fields to restore Weyl invariance lends support to our recent efforts to resolve and describe the Big Bang singularity \cite{Boyle:2018tzc, Boyle:2018rgh, Boyle:2021jej, Turok:2022fgq, Boyle:2022lcq, Boyle:2022lyw}.  When the trace of the stress tensor is zero, the Friedmann equation implies that the scale factor of the universe vanishes linearly in conformal time. Hence the big bang is an analytic, conformal zero of the metric and one can impose Dirichlet or Neumann boundary conditions on all fields there, in a manner which respects CPT and CT symmetry~\cite{Boyle:2018tzc, Boyle:2021jej,Boyle:2022lyw}. Second, the uniqueness of the vacuum state of the dimension-zero scalars provides a strong basis for a theory of primordial perturbations that is simpler, more predictive and less adjustable than inflation~\cite{perts}.

Understandably, we find this idea very intriguing.  But let us end by emphasizing some important unanswered questions:
\begin{itemize}
\item As mentioned above, the cancellation of all these gravitational anomalies requires that the Higgs field is composite or emergent.  Can this work in a way that agrees with the observed properties of the Higgs field and electroweak symmetry breaking?  

As pointed out by Bogoliubov {\it et al.} (see Ch. 10, p. 432 of Ref.~\cite{Bogoliubov}), fields of nonzero scaling dimensions can be obtained by exponentiating free dimension zero fields. For example, the operator $e^{i g \varphi}$, with $\varphi$ normalized as above, has a mass scaling dimension of $g^2/(16 \pi^2)$. This four dimensional mechanism is the analog of that used in string theory to build dimensionful vertex operators by exponentiating dimensionless world sheet coordinates (see {\it e.g.}, Ref.~\cite{Green:1987sp}). An enticing possibility is that an interacting, dimension one Higgs scalar doublet can be built in this way from dimension zero fields and other standard model operators.  (For other related ideas, see Ref.~\cite{Miller:2022qil}.)
 
\item Why 36? 

As yet, we have no conclusive answer. However, we hope the reader will indulge us in a little numerology, which may be suggestive. From the particle physics viewpoint, dimension zero scalars can be inserted as a ``phase" for every Weyl fermion, $\psi_a\rightarrow e^{i \varphi_a} \psi_a$, $a=1,\dots 48$. The theory would then be invariant under shifts $\varphi_a\rightarrow \varphi_a+\chi_a$ provided $\psi_a\rightarrow e^{-i\chi_a} \psi_a$.  This is in fact the ``gradient model" (for a single fermion) of Bogoliubov {\it et al.}~\cite{Bogoliubov}. Applying this to the Dirac Lagrangian for standard model fermions, $SU(3)\times SU(2)\times U(1)$ gauge transformations allow us to remove 12 phases, at linear order, so only 36 gauge invariant, dimension zero scalars remain. Adding the action (\ref{S_4_flat}) for each of these, thus making them dynamical, produces an interesting model. Similarly, when gravity is formulated as a gauge theory, a basic quantity is the field strength associated with the spin connection. This is a spacetime two form, in the adjoint representation of $SO(3,1)$ or $SO(4)$. As such, it has 36 components. For nice reviews, see Refs.~\cite{Celada:2016jdt,Krasnov:2020lku}.

\item We have only considered quantum fields in a fixed background spacetime (or, more correctly, a fixed conformal/Weyl class of background spacetimes). Including dynamical spacetime is essential and, most likely, will require going beyond the Euclidean approach employed here. 

\end{itemize}

{\bf Acknowledgements.}  We thank Sam Bateman, Paul Davies, Luigi Del Debbio, Michael Duff, Kurt Hinterbichler, Arkady Tseytlin, Grigory Volovik, and Roman Zwicky for helpful discussions and feedback.  Research at Perimeter Institute is supported by the Government of Canada, through Innovation, Science and Economic Development, Canada and by the Province of Ontario through the Ministry of Research, Innovation and Science.  The work of NT is supported by the STFC Consolidated Grant `Particle Physics at the Higgs Centre' and by the Higgs Chair of Theoretical Physics at the University of Edinburgh.

\end{document}